\documentclass[mathleft,fleqn,%
]{an}
\usepackage{graphicx}
\usepackage[varg]{txfonts}
\usepackage{natbib}
\bibpunct{(}{)}{;}{a}{}{,}
\begin{document}

\Pagespan{1}{}
\Yearpublication{2017}%
\Yearsubmission{2017}%
\Month{0}%
\Volume{999}%
\Issue{0}%
\DOI{asna.201400000}%

\title{Considerations on how to investigate planes of satellite galaxies}

\author{Marcel S. Pawlowski\inst{1}\fnmsep\thanks{Hubble Fellow, email:
        marcel.pawlowski@uci.edu}
\and J\"org Dabringhausen\inst{2}
\and Benoit Famaey\inst{3}
\and Hector Flores\inst{4}
\and Fran\c{c}ois Hammer\inst{4}
\and Gerhard Hensler\inst{5}
\and Rodrigo A. Ibata\inst{3}
\and Pavel Kroupa\inst{2,6}
\and Geraint F. Lewis\inst{7}
\and Noam I. Libeskind\inst{8}
\and Stacy S. McGaugh\inst{9}
\and David Merritt\inst{10}
\and Mathieu Puech\inst{4}
\and Yanbin Yang\inst{4}
}

\titlerunning{Considerations on how to investigate planes of satellite galaxies}

\authorrunning{Pawlowski et al.}
\institute{
Department of Physics and Astronomy, University of California, Irvine, CA 92697, USA
\and 
Charles University in Prague, Faculty of Mathematics and Physics, Astronomical Institute,
 V Hole\v{s}ovi\v{c}k\'ach 2, CZ-180 00 Praha 8, Czech Republic
\and 
Universit\'e de Strasbourg, CNRS, Observatoire astronomique de Strasbourg, UMR 7550, 
 11 rue de l'Universit\'e, F-67000 Strasbourg, France
\and 
GEPI, Observatoire de Paris, PSL Research University, CNRS, Univ. Paris Diderot, 
Sorbonne Paris Cité, Place Jules Janssen, F-92195 Meudon, France
\and 
University of Vienna, Department of Astrophysics, Tuerkenschanzstr. 17, 1180 Vienna, Austria
\and
Helmholtz-Institut f\"ur Strahlen- und Kernphysik, Rheinische Friedrich-Wilhelms-Universit\"at Bonn,
 Nussallee 14-16, D-53115 Bonn, Germany
\and 
Sydney Institute for Astronomy, School of Physics, A28, The University of Sydney, NSW 2006, Australia
\and 
Leibniz-Institut für Astrophysik, Potsdam, An der Sternwarte 16, D-14482 Potsdam, Germany
\and 
Department of Astronomy, Case Western Reserve University, 10900 Euclid Avenue, Cleveland, OH, 44106, USA
\and 
School of Physics and Astronomy and Center for Computational Relativity and Gravitation, Rochester Institute of Technology,
84 Lomb Memorial Drive, Rochester, NY 14623, USA
}

\received{XXXX}
\accepted{XXXX}
\publonline{XXXX}


\abstract{
The  existence of a spatially thin, kinematically coherent Disk of Satellites (DoS) around the Milky Way (MW), is a problem that often garners vivacious debate in the literature or at scientific meetings. One of the most recent incarnations of this wrangle occurred with two papers by Maji et al, who argued that these structures ``maybe a misinterpretation of the data''. These claims are in stark contrast to previous works. Motivated by this and other recent publications on this problem, we discuss necessary considerations to make, observational effects to consider, and pitfalls to avoid when investigating satellite galaxy planes such as the MW's DoS. In particular, we emphasize that conclusions need to have a statistical basis including a determination of the significance of satellite alignments, observational biases must not be ignored, and measurement errors (e.g. for proper motions) need to be considered. We discuss general problems faced by attempts to determine the dynamical stability of the DoS via orbit integrations of MW satellite galaxies, and demonstrate that to interpret simulations, it is helpful to compare them with a null case of isotropically distributed satellite positions and velocities. Based on these criteria, we find that the conclusions of Maji et al. do not hold up to scrutiny, and that their hydrodynamic cosmological simulation of a single host shows no evidence for a significant kinematic coherence among the simulated satellite galaxies, in contrast to the observed MW system.}

\maketitle

\section{Introduction}
\label{sect:introduction}

\citet{LyndenBell1976} and \citet{KunkelDemers1976} first noted that the Milky Way (MW) satellite galaxies LMC, SMC, Draco, Ursa Minor, Sculptor and Fornax (\citealt{KunkelDemers1976} also included Leo I and II)\footnote{Carina \citep{Cannon1977}, Sextans \citep{Irwin1990}, and Sagittarius \citep{Ibata1994} were not known at the time of the work of \citet{LyndenBell1976} and \citet{KunkelDemers1976}, contrary to the attribution to them of 11 satellites by e.g. \citet{Sawala2014}, \citet{Cautun2015}, \citet{Maji2017a}, \citet{Maji2017b}.}, as well as several star clusters, align along a great polar circle around the MW. \citet{Kroupa2005} showed that the then-known eleven MW satellite galaxies form a Disk of Satellites (DoS): their distribution is highly flattened and oriented almost perpendicular to the MW. They also pointed out that this is in contrast to predictions from the standard cosmological model.

While the existence of planes of satellite galaxies and signatures of their kinematic coherence have now been firmly established for the MW and M31 systems \citep{Pawlowski2012,Conn2013,Ibata2013,Pawlowski2016}, and indications for such structures have been found beyond the Local Group \citep{Ibata2014,Tully2015,Mueller2016}, the origin, stability, and most importantly interpretation of the satellite planes within the context of cosmological models is still debated \citep[e.g.][]{Libeskind2009,Kroupa2010,Hammer2013,Pawlowski2014,Yang2014,Cautun2015,Libeskind2015,Smith2016,Angus2016}.

\citet{Maji2017a,Maji2017b} recently joined this discussion with two papers. In \citet{Maji2017a}, the authors call into question the evidence for the MW DoS and its kinematic coherence, suggesting that it might be a ``misinterpretation of the data''. In \citet{Maji2017b} they then demonstrate that their single hydrodynamical simulation of a MW-like host halo does not contain a strongly flattened or kinematically coherent satellite plane.

Given our previous experience in studying the phase-space distribution of the MW satellite galaxies, we feel obliged to comment on these contributions due to their combination of far-reaching claims and the lack of supporting evidence. Their claims are plagued with methodological problems; inadequate statistical analysis; and other inaccuracies of varying degrees of severity. These shortcomings have motivated us to present some general considerations on investigations of satellite galaxy planes.

We consider the following elements to be required in order to come to meaningful conclusions on the existence and prominence of the DoS and kinematic correlations among simulated satellite galaxies. The analysis should be statistically sound, and thus transparently measure the significance of its results. It should take (at least the dominant) measurement errors into consideration, as well as be aware of the influence of observational biases. Finally, to be able to determine the strength of a phase-space coherence requires comparison with a well defined null hypothesis, preferably an isotropic model (to which the before mentioned biases should be applied as well). 

We will discuss these aspects in more detail in the following and assess whether they have been sufficiently considered in recent works, with an emphasis on those making the strongest claims \citep{Maji2017a,Maji2017b}. To facilitate easier comparison of this comment with their papers, we use the term DoS here, even though Vast Polar Structure (VPOS) would be more accurate since it implies the alignment of additional objects and streams found by \citet{Pawlowski2012}.

\section{Statistical Significance and Measurement Errors}
\label{subsect:significance}

\begin{figure}
   \centering
   \includegraphics[width=\linewidth]{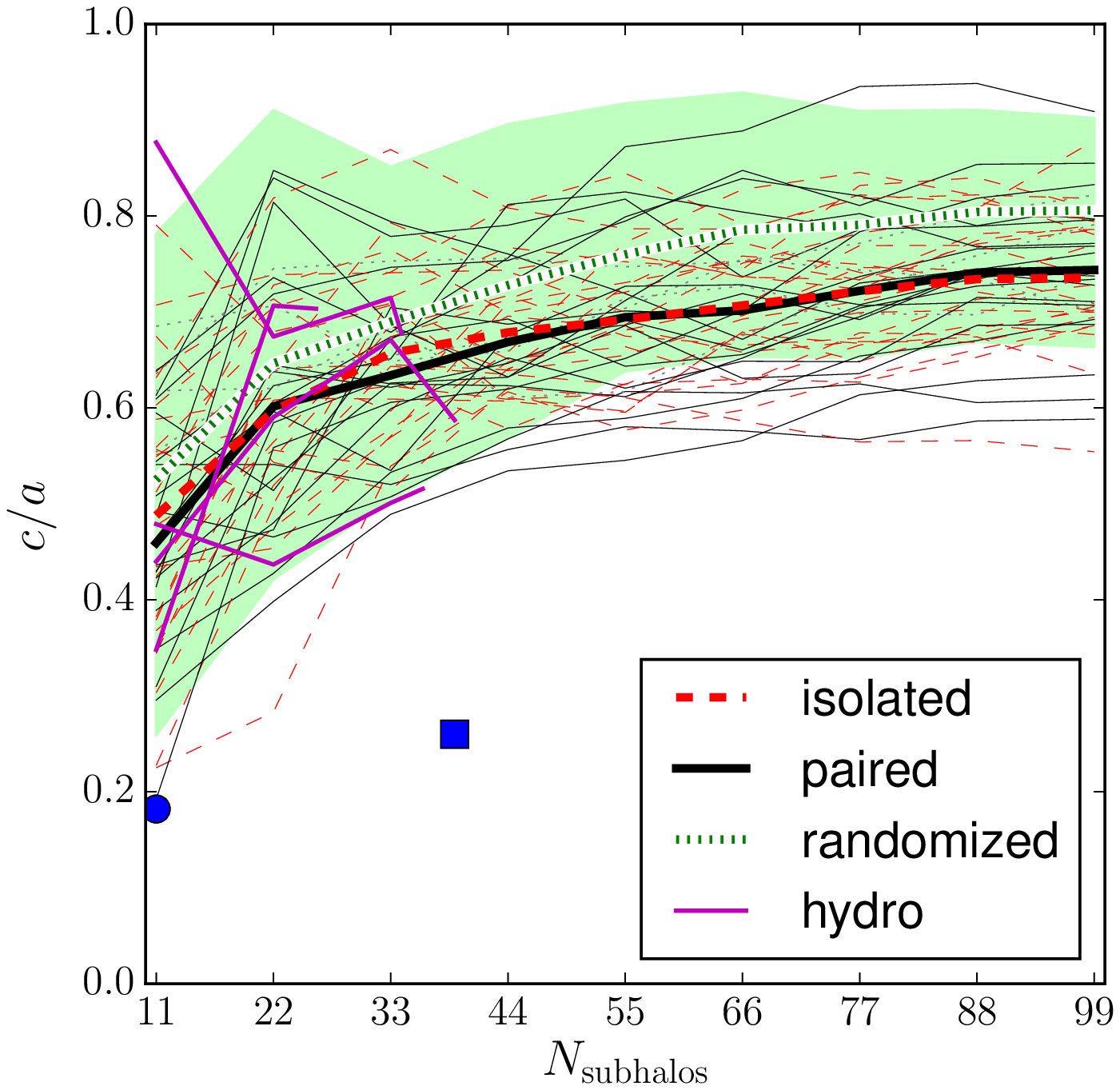}
   \caption{
The minor-to-major axis ratio flattening of sub-halo systems in the ELVIS suite of cosmological simulations of MW and Andromeda galaxy equivalents \citep{GarrisonKimmel2014}, and the satellite galaxy systems in the four hydrodynamical cosmological simulations used in \citet{Ahmed2017}.
For larger numbers of considered sub-halos (ranked by their abundance-matching based luminosity) or satellite galaxies (ranked by their stellar mass), the overall flattening declines with increasing sample size. The 11 classical MW satellites nevertheless have a stronger flattening than the simulations ($c/a = 0.18$, blue dot). \citet{Maji2017a} report $c/a \sim 0.26$\ for a sample of 39 MW satellites (blue square). While they argue that such an increase shows that the distribution becomes significantly more isotropic, this figure demonstrates that a slightly increased $c/a$\ for a larger sample size might in fact have a higher significance of the flattening compared to isotropic (''randomized'') and simulated distributions.
See also figure 3 in \citet{PawlowskiMcGaugh2014}.
   }
              \label{fig:axisratios}
\end{figure}

\begin{figure}
   \centering
   \includegraphics[width=\linewidth]{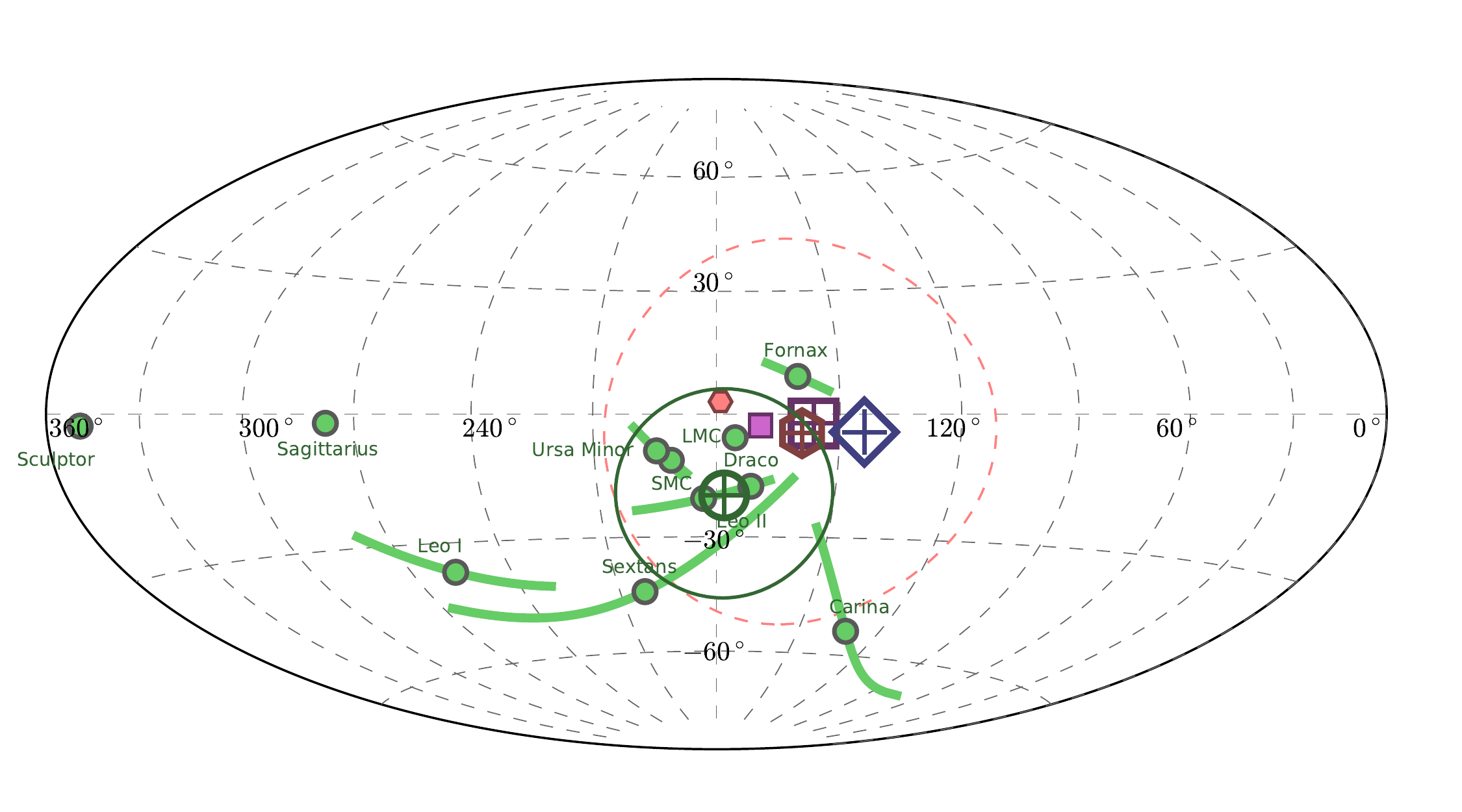}
   \caption{
Distribution of the orbital poles of the classical MW satellite galaxies, updated with recent HST proper motion measurements \citep{Pryor2015, Piatek2016, Sohn2017}. The green great-circle segments indicate the one-sigma uncertainties in the directions of the orbital poles due to the proper motion measurement uncertainties. The magenta square with a plus sign indicates the direction of the normal to the plane fitted to the positions of the 11 classical satellite galaxies. Within their uncertainties, eight of the 11 satellites are consistent with co-orbiting in the DoS, while another one (Sculptor) is counter-orbiting. Only Leo I (which might not be bound to the MW) and Sagittarius (which due to its position can not possibly orbit in the DoS) are inconsistent with co-orbiting by more than one sigma.
   }
              \label{fig:orbitalpoles}
\end{figure}

To quantify whether an apparent correlation is real, it is necessary to determine its statistical significance. In doing so, one can estimate how likely it is that the observed configuration could arise from random chance. When working with phase-space distributions of satellite galaxy systems, it has proven useful to compare their properties with those drawn from isotropic distributions (in position and/or velocity). This allows for an efficient and easily implemented test of significance, which has the benefit of being agnostic to underlying cosmological models and thus it is applicable to observations and simulations alike.

Observational uncertainties of measured parameters need to be taken into account in scientific analyses, especially when investigating potentially correlated systems. In such cases, measurement uncertainties can constitute an important contribution to the measured scatter in relations, or even dominate entirely in case of negligible intrinsic scatter (see e.g. \citealt{Lelli2016} for a recent example of measured vs. intrinsic scatter).

\citet{Maji2017a} fit planes to, and measure the corresponding flattening of, three samples of MW satellite galaxies: the 11 classical ones, as well as samples of 27 and 39 satellites. Information on which objects constitute the larger samples, and what positions and distances were adopted for them, has not been provided. \citet{Maji2017a} report both the minor-to-major axis ratios $c/a$\ of these distributions as well as the absolute root-mean-square heights. We will focus on the former in the following, but similar arguments apply to the latter. \citet{Maji2017a,Maji2017b} consider three cases for weighting the distance of the individual galaxies in the fit of a plane to the DoS and the estimate of $c/a$\ of the DoS, namely no weighting, a down-weighting with $1/r$, and a down-weighting with $1/{r^2}$\ (where $r$ is the Galactocentric distance of a satellite). Reasons for why such a weighting of satellites in the plane fitting routine is less informative have been discussed at length in \citet{Pawlowski2015}. 
Basically, down-weighting by radius has the effect of suppressing the information content of the observed satellite positions. In particular, a weighting by $1/{r}$\ effectively projects all positions onto a unit-sphere before the flattening of the distribution is calculated. This has the consequence that only the angular alignment of the satellites is considered, and that available information is ignored because the dimensionality of the data is reduced to only two out of three dimensions. While downweighting results in slightly reduced sensitivity to distant outliers, it leads to increased sensitivity to angular outliers at small Galactocentric distances which are expected for a structure of non-zero width. Furthermore, \citet{Pawlowski2015} showed that a test based on downweighting is less discriminating and biases towards finding an agreement with the observed flattening of the MW satellite distribution. Due to these arguments, we discuss in the following only the case where the satellite galaxies are not weighted by their distance.

The axis ratio $c/a$\ increases from 0.18 for the 11 classical satellites, to about 0.26 for the sample of 39 satellites in \citet{Maji2017a}. They argue that such an increase demonstrates that the DoS becomes ``significantly more isotropic and thicker''. However, no significance has been measured by them, such that this judgement seems questionable. This worry is corroborated by Fig. \ref{fig:axisratios}, which shows how the axis ratio flattening of simulated satellite distributions changes with the number of considered objects. The data in this plot is from the ELVIS suite of standard cosmology simulations, and the effect of increasing satellite number on $c/a$\ and other parameters has already been discussed in \citet{PawlowskiMcGaugh2014}. In summary, for low numbers of satellites the true thickness of the overall satellite distribution is underestimated: fitting a plane to a smaller number of satellites will on average result in a thinner plane height and a smaller axis ratio. This is because the few most distant satellites will define the plane fit and thus have small offsets, while those satellites at small radii will have small distances from any plane. In the extreme case of three satellites, the best-fit plane would have zero thickness. Only once the number of satellites increases, the measured thickness starts to approach the true flattening of the overall distribution. This also explains why the discrepancy between simulated and randomized (isotropic) satellite distributions are more pronounced for larger numbers of considered satellites.

Figure \ref{fig:axisratios} shows that the value of $c/a$\ is expected to increase rapidly for larger sample sizes (in standard cosmological models as well as isotropic systems), such that even a moderate increase in the observed $c/a$\ to 0.26 at 39 satellites in fact \textit{increases} the distance between the observed and the expected $c/a$, both for simulated satellite systems (black solid and red dashed lines) as well as for isotropic distributions (green dotted line and green shaded area). Had \citet{Maji2017a} performed a comparison with such expectations and thereby determined the statistical significance of the flattening, they might well have found the significance to in fact increase for the larger sample sizes. The data might thus point to the opposite of their interpretation: that the increase in sample size strengthens the evidence for the DoS, rather than diminishes it.

The inclusion of baryonic physics can affect the distribution of satellite galaxies in simulation. In particular, it has been found that satellite systems in hydrodynamical simulations are more radially extended due to the stronger tidal disruption close to the central galaxy \citep{Ahmed2017,GarrisonKimmel2017}. Furthermore, even among dark-matter-only and hydrodynamical simulations based on the same initial conditions, different sets of satellites might be identified as the most luminous ones. Despite these issues, we find that the effect of rising axis ratios for increasing satellite numbers also tends to be present in hydrodynamical simulations, as exemplified by the magenta lines in Fig. \ref{fig:axisratios}. These are based on the four simulations used in \citet{Ahmed2017}, with the satellites ranked according to their stellar mass. Due to limited resolution, the lines are only plotted up to the maximum number of satellites that contain at least three star particles, follwing the choice made in \citet{Ahmed2017}. As for the dark-matter-only simulations, all four of these satellite systems are considerably less flattened than the observed VPOS.

In their comparison to a single hydrodynamical simulation, \citet{Maji2017b} find $c/a \approx 0.3$\ for the 11 most massive simulated dwarf galaxies, which they argue is close to the observed value of 0.18. However, to be able to meaningfully interpret differences in numerical values, a statistical analysis of the distribution of c/a values is required. Only this can inform judgement of whether $c/a \approx 0.2$\ and $c/a \approx 0.3$\ are close, or in significant conflict. As can be seen in Fig. \ref{fig:axisratios}, $c/a$\ values of ~ 0.3 can be reached by the shown dark matter only simulations, whereas the observed value of 0.18 is so low that it is never reached, which hints at the latter option. The model by \citep{Maji2017b} may produce a different result, but this can only be determined if they run their simulations according to their model sufficiently often to enable a statistical analysis.

Similar arguments apply to investigations of the distribution of orbital poles of the MW satellite galaxies. In fig. 2 of \citet{Maji2017a}, they use data compiled in \citet{PawlowskiKroupa2013} to largely reproduce the latter study's figure 1, in particular the distribution of the orbital poles of the MW satellite galaxies. Additional HST measurements of proper motions \citep{Pryor2015, Piatek2016} -- which result in updated, more tightly clustered orbital poles -- have not been considered by \citet{Maji2017a}. In Fig. \ref{fig:orbitalpoles} we show an updated version of the plot originally published in \citealt{PawlowskiKroupa2013}. Furthermore, the orbital pole uncertainties (green great-circle segments in Fig. \ref{fig:orbitalpoles}) have been omitted in \citet{Maji2017a}. The authors then set an orbital pole alignment criterion, according to which an orbital pole has to be within $45^\circ$\ of the normal vector to the best-fit plane of satellites. Six of the 11 classical MW satellite galaxies fall within this category around one of the two normal directions. Not mentioned by \citet{Maji2017a} is that Sculptor is also within $45^\circ$\ of the plane normal but in a counter-orbiting orientation, such that 7 out of 11 objects have orbital poles that align to within $45^\circ$ with the satellite plane normal.

\citet{Maji2017a} additionally neglect to mention that two other orbital poles are just barely excluded by the cut at $45^\circ$: Carina's most-likely orbital pole is at an angular distance of $45.7^\circ$, and Sextans' is at $49.8^\circ$\ \citep{PawlowskiKroupa2013}. Both have orbital pole uncertainties that extend well into the region close to the satellite plane normal. Within just their $1\sigma$\ uncertainties, Sextans is consistent with aligning perfectly with the satellite plane normal, and Carina aligns to within $15^\circ$. 

As was the case for the measure of the plane flattening, \citet{Maji2017a} do not attempt to determine the significance of the orbital  pole concentration, while this method would deliver more conclusive results than arguments based on an arbitrary $45^\circ$\ cut-off. In fact, it was shown in \citet{PawlowskiKroupa2013} that the probability to find the eight most-aligned orbital poles to cluster at least as tightly as observed it 0.1 per cent if the distribution of orbital poles was isotropic, and that the average direction of those orbital poles differs by only $18.9^\circ$\ from the normal vector of the best fitting plane to the DoS \citep[e.g.][]{PawlowskiMcGaugh2014}.

\citet{Maji2017a} therefore neither base their conclusion on a statistically sound determination of the orbital pole correlation's significance, nor do they take the sometimes substantial measurement errors affecting proper motions into account. Measurement errors were discussed in \citet{PawlowskiKroupa2013}, who also pointed out that orbital poles with smaller uncertainties align more closely with the satellite plane normal.

\section{Orbit Calculations and Measurement Errors}

Even if one grants that the DoS is currently a thin and coherent structure, one could still argue that this is a chance occurrence, and that the same satellites will exhibit a very different distribution in the future (and also in the past). Some recent papers attempt to explore this possibility by carrying out orbit integrations, advancing the positions of the satellites backward \citep{LipnickyChakrabarti2016} or forward in time \citep{Maji2017a}, starting from their currently-measured positions and velocities.

There is a problem with drawing conclusions from simulations like these. The hypothesis being tested is that the DoS is a long-lived structure; that is, that the orbits of the satellites are narrowly confined in integral-space. Suppose that the hypothesis is true. It follows that {\it any} errors in the {\it simulated} orbits -- due to errors in the initial positions or velocities, in the assumed form of the potential, etc. -- are {\it guaranteed} to result in a decrease in time of the system's coherence. Given that the current satellite distribution {\it is} highly coherent, a loss of coherence with time is almost certain to occur in simulations of this sort, and it can not be taken as evidence that the existing correlations are short-lived. In order to make such an argument, one would need to show that there is {\it no} set of initial conditions consistent with the observed proper motions and positions that can result in long-term coherence, and furthermore that this is true in every potential that is consistent with observations of the MW.

Calculating a satellite galaxy's orbit around the Milky Way requires precise knowledge of the underlying gravitational potential, as well as exact measurements of the satellite's current position, distance, line-of-sight velocity and proper motion, and of the Galactocentric distance of the Sun and the Solar motion with respect to the Galactic center. Many of these quantities are not well constrained. Even for a parameter as fundamental as the total mass of the MW, current estimates range from 0.6 to more than $2.0 \times 10^{12}\,\mathrm{M}_{\sun}$\ \citep{Watkins2010, Deason2012, Boylan-Kolchin2013, Gibbons2014}. Likewise, some studies found the shape of the MW halo to be close to spherical, while others prefer a triaxial halo \citep{Ibata2001,Fellhauer2006,LawMajewski2010,Kuepper2015}, which can have profound effects on orbital stability of a satellite plane \citep{Fernando2017}. The matter is further complicated by the fact that the satellites orbit in an evolving potential.

In order to disprove the hypothesis that the DoS is rotationally supported, it is therefore not sufficient to integrate the satellite orbits based on their most likely proper motions in a single fixed potential. Instead, the following, stronger claim must be proven: Even considering the substantial proper motion uncertainties (as well as other measurement uncertainties and the fact that the profile, shape, and depth of the MW potential is not well known), there is no configuration of possible satellite orbits that result in a stable DoS. \textit{ What must be shown is thus not that the most-likely proper motion measurements result in a stable satellite plane. Rather, it needs to be demonstrated that within the range of possible proper motions of all satellites, no mutual configuration exists which kept the objects currently orbiting within the DoS in a coherent structure in the past.}

The work of \citet{LipnickyChakrabarti2016} is based on integrating the orbits of the MW satellite galaxies backward in time. They find that a plane fitted to the backward-integrated satellite positions has an increased height, and conclude that the DoS is not a dynamically stable structure. However, they do not take the before discussed measurement uncertainties into account in the orbit calculations. Instead, they determine how long on average a perfect ring of satellite galaxies on circular orbits remains coherent if errors are added to the initial velocities. They define coherence as having a ratio of root-mean-square height $R_\mathrm{rms}$\ to median distance from the host $D_\mathrm{med}$\ of $\frac{R_\mathrm{rms}}{D_\mathrm{med}} < 0.33$. Since their test-case is a satellite ring of 150\,kpc radius, this implies that their toy model is allowed to increase in thickness from 0\,kpc (perfect initial alignment) to $\approx 50$\,kpc ($0.33 \times 150$\,kpc), whereas \textit{any} increase in thickness for the observed system is interpreted as a sign of dynamical instability.

Similarly, in their fig. 3, \citet{Maji2017a} show their results for the future orbits of those MW satellite galaxies with measured proper motions, apparently assuming (they do not specify this) their most-likely 3D velocities. They plot the satellites' current, as well as their future positions. Future positions after 1 and 5 Gyr are shown (as reported both in the text and the caption). These orbit integrations do not take the before mentioned sources of uncertainties into account. 

That \citet{LipnickyChakrabarti2016} and \citet{Maji2017a} find an apparent dispersion of the satellite plane is therefore not surprising, and it does not allow any meaningful conclusion to be drawn as to the dynamical state of the DoS. They did not test whether the observed proper motions are consistent \textit{within their uncertainties} with a dynamically stable satellite galaxy plane. We further note that either the reported times, the plotted distances, or the overall orbit integration in \citet{Maji2017a} is wrong: with typical velocities of 200\,km\,s$^{-1}$, a MW satellite would have moved by about 1\,Mpc in 5\,Gyr (and consequently many would have completed multiple orbits). This is substantially more than the few 10 kpc of motion shown in their figure.

\section{Observational Biases}
\label{sect:biases}

\begin{figure}
   \centering
   \includegraphics[width=\linewidth]{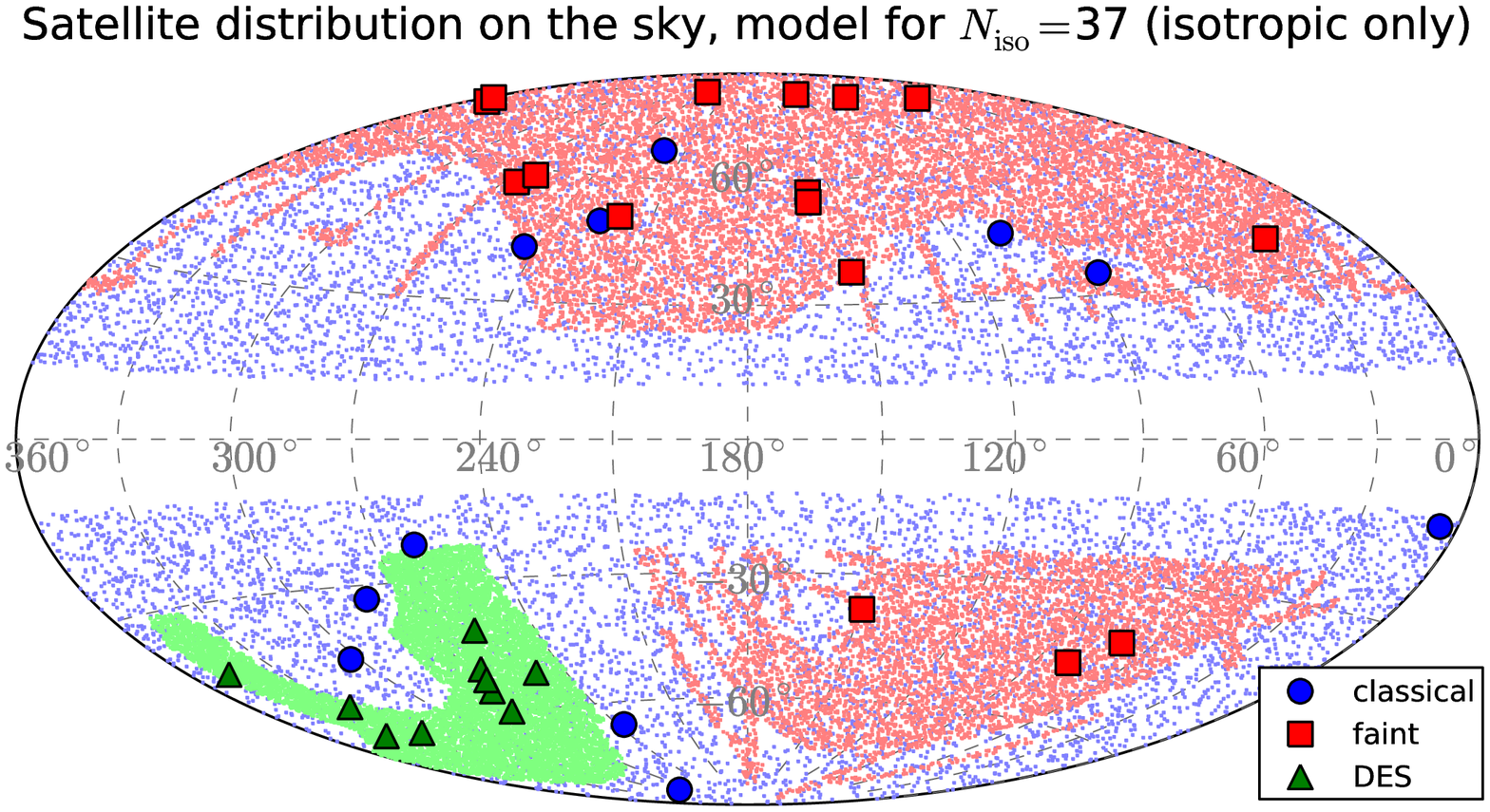}
   \caption{
Distribution of MW satellite galaxies (classical in blue, SDSS based discoveries in red, first year DES discoveries in green) and the corresponding survey footprints. For the classical satellites, it is assumed that the MW disc prohibits discovery within $b = \pm 12^\circ$. This illustrates that the spatial distribution of discoverable MW satellite galaxies suffers from complicated positional biases. An analysis expanding beyond the 11 classical satellite galaxies to fainter objects thus needs to consider and model these complex biases in order to arrive at a statistically sound comparison of the observed flattening of the satellite system and simulations. 
Figure adopted from fig. 1 in \citet{Pawlowski2016} after adding DES information to original.
   }
              \label{fig:biases}
\end{figure}

The observed distribution of satellite galaxies around the Milky Way is affected by a number of observational biases. Crowding by foreground stars and obscuration by MW dust make it very difficult to detect even bright satellites close to the galactic plane. Fainter objects can only be discovered in deep surveys, the two most successful of which have been the Sloan Digital Sky Survey (SDSS, \citealt{SDSSpaper}) and the Dark Energy Survey (DES, \citealt{DESpaper}). As shown in Fig. \ref{fig:biases}, they have been confined to specific regions of the sky. Faint dwarf galaxies might still be lurking in the regions not observed. The survey footprints can thus introduce strong spatial biases in the distribution of known MW satellite galaxies, which are further complicated by other effects affecting dwarf galaxy discoverability, such as local Galactic foreground and variations of observing conditions between different survey pointings. Modelling such spatial biases is particularly important when investigating the spatial distribution of satellite galaxies. It has been demonstrated that approximate approaches to such modelling are possible and can be applied to statistical analyses of the alignment of MW satellite galaxies \citep[e.g.][]{Metz2009,Pawlowski2016}.

Furthermore, the known sample of faint dwarf galaxies is radially biased. Due to such galaxies' very low luminosity and surface brightness, current surveys can only discover these objects nearby, instead of within the whole MW halo \citep{Tollerud2008}. The inclusion of faint MW satellite galaxies in a measure of the satellite plane flattening thus introduces biases whose effect on the resulting values for axis ratios and plane height are difficult to evaluate, questioning their use as reliable probes of the overall phase-space correlation of the MW satellite system.

When \citet{Maji2017a} measure the spatial flattening of the MW satellite system, they do not attempt to consider the effects of MW disk obscuration, survey footprints, or radial biasing of faint satellite galaxies. As such, it is therefore more than questionable whether the sheer increase in the number of tracers used to measure the satellite system's flattening actually provides more information on the satellite distribution than on the underlying survey strategies.

The points on survey footprints and radial biases not only apply to determinations of the significance of the flattening of an observed satellite system, but also to comparisons of observed systems with simulations. As has been demonstrated in \citet{Pawlowski2014}, the size and shape of the considered volume can significantly affect results on the frequency of narrow satellite galaxy planes. For example, more narrow volumes bias towards more narrow satellite distributions.

Survey shapes and selection effects have not been considered in numerous studies, in particular those selecting satellites from the entire virial volume of a simulated host halo \citep[e.g.][]{Sawala2014,Buck2015,Ahmed2017}. In \citet{Maji2017b}, the authors not only ignore survey footprints and radial biases when comparing observations with their single simulation, the compared volumes are substantially different, too. Large portions of the \citet{Maji2017b} analysis focus on a volume of 1\,Mpc in radius\footnote{One might suspect from their fig. 1 that the considered volume is either $1\,\mathrm{Mpc}/h$\ (with $h \approx 0.7$), or that the virial radius of the host galaxy is smaller than reported. In the figure, the circle indicating the host's virial radius has a radius of about one third of the distance to the 0.5 Mpc tick mark, which would correspond to a virial radius of 167\,kpc instead of the reported $\approx 240\,\mathrm{kpc}$. This is curiously close to $0.7 \times 240\,\mathrm{kpc} = 168\,\mathrm{kpc}$.}, while \citet{Maji2017a} concentrates on the virial radius of the MW, which for their simulation they give as 240\,kpc. Thus the compared volumes are almost two orders of magnitude different.
In this regard, the two volumes thus trace very different regimes, for example it is no surprise that at 1\,Mpc the existence of filamentary structures result in smaller c/a values. Differences between the simulation and the observed MW satellite distribution are thus less dominated by sample sizes, but rather by probing very different structures. 
When they restrict the satellite sample to smaller numbers and the virial volume, \citet{Maji2017b} find less narrow distributions than observed, in line with previous results based on dark matter only simulations \citep[e.g.][]{PawlowskiMcGaugh2014}.

\section{An Isotropic Comparison Model}

As discussed in Sect. \ref{subsect:significance}, it can be very instructive to use isotropic distributions as a benchmark to develop an expectation that allows one to identify if simulations contain new correlations or behaviour. A benefit of isotropic models is that they do not display any \textit{intrinsic} flattening or orientation, which for example allows one to study what effect mere low-number-statistics has on the measured flattening of satellite samples (e.g. dotted line and green region in Fig. \ref{fig:axisratios}), or how survey footprint shapes affect the expected flattening of a satellite system \citep{Pawlowski2016}. From a practical standpoint, one benefit is that expectations for isotropic distributions can be derived from basic geometry on a sphere. Oftentimes, expected fractions of aligned satellites are easily determined by calculating the corresponding area on a sphere. A useful application is the quick assessment of kinematic alignments, addressing questions such as whether more satellite galaxies move close to (or orbit within) a common plane than expected for an isotropic distribution.

As an example, we will in the following use some geometric considerations and the information \citet{Maji2017b} provide about their simulation to make such a comparison. This addresses the lack of any assessment of the degree of the kinematic alignments presented by \citet{Maji2017b}, which made it difficult to judge whether the reported fractions of kinematically aligned dwarf galaxies are indicative of additional coherence beyond mere isotropy.

\citet{Maji2017b} define a dwarf galaxy as moving in the plane of satellites if its 3D velocity vector aligns to within $45^\circ$\ or less with the plane. According to this criterion, a satellite on an orbit perpendicular to the plane and currently at its most distant point from the plane (thus currently moving in a direction parallel with the plane) is kinematically aligned. This criterion thus considers only the 3-D velocities, while it completely ignores the positions, which makes it so weak that the vast majority of the satellites fulfil it even in the case of complete isotropy.

\citet{Maji2017b} report that 77 of the 106 luminous dwarfs in their simulation (using all dwarfs out to 1\,Mpc) fulfil their kinematic alignment criterion. Note, however, that the histogram in their figure 7, while referred to as showing that 77 out of 106 simulated dwarfs have velocities within $45^\circ$\ of the best-fit plane, only contains a total of 77 data points, not 106.
If we nevertheless assume that the text is correct, this results in 73 per cent alignment. While \citet{Maji2017b} do not provide any means to judge whether this is a large, small, or modest alignment, we can calculate the expected fraction for an isotropic distribution of velocity vectors. The fraction of isotropic directions which align to within an angle $\theta$\ with any given plane is $\sin{\theta}$. Thus, for $\theta = 45^\circ$\ we expect 71 per cent of the vectors to align\footnote{Note that because the plane to which the velocities are compared in \citet{Maji2017b} is defined by the simulated dwarf galaxy positions, one can expect a modest increase in this fraction.}. \textit{This is fully consistent with the fraction found by \citet{Maji2017b}, indicating that this measure reveals no evidence that the simulation results in an increased kinematic alignment of dwarf galaxies compared to an isotropic model}\footnote{If we instead read off the numbers from the histogram in their fig. 7, there are 21 satellites with velocities inclined by more than $45^\circ$\ from the best-fit plane, vs. 56 with more closely aligned velocity vectors. These numbers again give 73 per cent of satellites with velocities within $45^\circ$\ of the best-fit plane, which again is fully consistent with the expected fraction from an isotropic sample.}.

\citet{Maji2017b} continue by counting the number of simulated satellites which have an orbital pole aligned to $45^\circ$\ or better with the normal vector of the plane fitted to the satellite positions. Out of the 77 satellites they find 18 to be co- and 19 to be counter-rotating\footnote{This is a non-standard interpretation of counter-rotating. The term usually refers to the opposite of the majority direction, in which case the number of counter-rotating objects cannot exceed the number of co-rotating ones.}. For an isotropic distribution of orbital poles, a fraction of $1 - \cos{\tau}$\ of poles will align to an angle better than $\tau$\ with the normal vector of any given plane. For $\tau = 45^\circ$, this corresponds to 29 per cent. \citet{Maji2017b} report 18 + 19 = 37 out of a total of 77, which is 48 per cent and thus appears high. However, the 77 are allegedly selected out of a total of 106 for having velocity vectors within $45^\circ$\ of the satellite plane. Conversely, this implies that the remaining 106 - 77 = 29 can not have an orbital pole closer than $45^\circ$\ to the satellite plane normal: a dwarf galaxy's orbital pole is always perpendicular to the velocity vector, if the latter has an angle of more than $45^\circ$\ from a plane, every vector perpendicular to it will always be within $45^\circ$\ of the plane, or more than $45^\circ$\ from the plane normal. Thus, the fraction to compare to an isotropic model is 37/106 (35 per cent), which is only slightly higher than the isotropic expectation. The offset is in the direction expected for the position-dependent plane orientation, and within the expected variation from standard poisson errors ($\sqrt{37} \approx \pm 6$, corresponding to a range in the fraction from 29 to 41 per cent.).

It is interesting to note that the observed MW system is inconsistent with these numbers, with at least six (possibly eight within their uncertainties, see Sect. \ref{subsect:significance}) of the 11 classical satellites having co-orbiting orbital poles within $45^\circ$\ of the DoS plane normal, plus one counter-orbiting (i.e. 64 to 81 per cent fulfilling their criterion). Nevertheless, \citet{Maji2017b} then seem to argue that the lack of a kinematic coherence in their simulation suggests that the observed DoS does not have a coherent rotation. This is a false conclusion, because the results of a simulation can not trump existing observational data when it comes to the nature of a system that is observed in reality.

Finally, it should be emphasized that the simulation results of \citet{Maji2017b} are based on a single host halo and therefore lack statistical significance. Furthermore, no information has been provided on whether that halo's central galaxy, its environment or local cosmography have any resemblance with the MW (e.g. motion along a filament towards Virgo, evacuation from the Local Void, group configuration with M31).

\section{Conclusion}
\label{sect:conclusion}

We have discussed several effects that need to be considered when investigating satellite galaxy planes like the DoS. In particular, conclusions need to have a statistical basis which mandates that the significance of satellite alignments be determined. Observational biases can strongly affect the spatial distribution of satellite galaxies, and must not be ignored in comparisons to models or simulations. Measurement errors such as for proper motions need to be considered, especially since they affect any attempts to determine the dynamical stability of the DoS via orbit integrations. Showing that the most-likely proper motions results in a dispersing satellite plane is not sufficient to rule out the hypothesis that the DoS is a long-lived structure, because if it is kinematically coherent any errors in the simulated orbits are guaranteed to falsely decrease the system's coherence over time. We have also argued that it is helpful to interpret simulation results by comparing them with an alternative model of isotropically distributed satellite velocities.

Applied to the specific examples that motivated this work, we find that the contributions by \citet{Maji2017a,Maji2017b} lack the necessary analyses to come to meaningful conclusions on the existence of the DoS and on its kinematic correlation. A statistical foundation and measurements of the significance of presented results are missing in their contributions. Had these been carried out, they would in all likelihood have pointed to the opposite of the authors' claims. The measurement errors for the MW satellite galaxy proper motions, as well as the uncertain nature of the MW potential, have not been taken into consideration. The effect of satellite sample size on the flattening of a satellite system is not a novel finding but has been demonstrated and discussed in past publications \citep[e.g.][]{Wang2013,PawlowskiMcGaugh2014}, and it does not imply that the DoS might be an effect of small sample size. The influence of observational biases was ignored, and the structure of MW satellite galaxies was inappropriately compared to that of simulated dwarf galaxies drawn from a much larger volume. We therefore strongly reject the suggestion that the observed DoS might be ``a misinterpretation of the data''.


\acknowledgements
We would like to thank Alyson Brooks for providing data for the four hydrodynamical simulations analysed in \citet{Ahmed2017}. M. S. P. acknowledges that support for this work was provided by NASA through Hubble Fellowship grant \#HST-HF2-51379.001-A awarded by the Space Telescope Science Institute, which is operated by the Association of Universities  for  Research  in  Astronomy,  Inc.,  for  NASA,  under  contract  NAS5-26555. G. F. L. thanks the Australian Research Council for support through his Future Fellowship (FT100100268) and Discovery Project (DP110100678). D. M. was supported by the National Science Foundation under grant no. AST 1211602 and by the National Aeronautics and Space Administration under grant no. NNX13AG92G.

\clearpage


\begin{thebibliography}{}

\bibitem[Ahmed et al.(2017)]{Ahmed2017} Ahmed, S.~H., Brooks, A.~M., \& Christensen, C.~R.\ 2017, \mnras, 466, 3119 

\bibitem[Angus et al.(2016)]{Angus2016} Angus, G.~W., Coppin, P., Gentile, G., \& Diaferio, A.\ 2016, \mnras, 462, 3221 

\bibitem[Boylan-Kolchin et al.(2013)]{Boylan-Kolchin2013} Boylan-Kolchin, M., Bullock, J.~S., Sohn, S.~T., Besla, G., \& van der Marel, R.~P.\ 2013, \apj, 768, 140 

\bibitem[Buck et al.(2015)]{Buck2015} Buck, T., Macci{\`o}, A.~V., \& Dutton, A.~A.\ 2015, \apj, 809, 49 

\bibitem[Cannon et al.(1977)]{Cannon1977} Cannon, R.~D., Hawarden, T.~G., \& Tritton, S.~B.\ 1977, \mnras, 180, 81P 

\bibitem[Cautun et al.(2015)]{Cautun2015} Cautun, M., Bose, S., Frenk, C.~S., et al.\ 2015, \mnras, 452, 3838 

\bibitem[Conn et al.(2013)]{Conn2013} Conn, A.~R., Lewis, G.~F., Ibata, R.~A., et al.\ 2013, \apj, 766, 120 

\bibitem[Deason et al.(2012)]{Deason2012} Deason, A.~J., Belokurov, V., Evans, N.~W., et al.\ 2012, \mnras, 425, 2840 

\bibitem[Gibbons et al.(2014)]{Gibbons2014} Gibbons, S.~L.~J., Belokurov, V., \& Evans, N.~W.\ 2014, \mnras, 445, 3788 

\bibitem[Garrison-Kimmel et al.(2014)]{GarrisonKimmel2014} Garrison-Kimmel, S., Boylan-Kolchin, M., Bullock, J.~S., \& Lee, K.\ 2014, \mnras, 438, 2578 

\bibitem[Garrison-Kimmel et al.(2017)]{GarrisonKimmel2017} Garrison-Kimmel, S., Wetzel, A.~R., Bullock, J.~S., et al.\ 2017, arXiv:1701.03792 

\bibitem[Fellhauer et al.(2006)]{Fellhauer2006} Fellhauer, M., Belokurov, V., Evans, N.~W., et al.\ 2006, \apj, 651, 167 

\bibitem[Fernando et al.(2017)]{Fernando2017} Fernando, N., Arias, V., Guglielmo, M., et al.\ 2017, \mnras, 465, 641 

\bibitem[Hammer et al.(2013)]{Hammer2013} Hammer, F., Yang, Y., Fouquet, S., et al.\ 2013, \mnras, 431, 3543 

\bibitem[Ibata et al.(1994)]{Ibata1994} Ibata, R.~A., Gilmore, G., \& Irwin, M.~J.\ 1994, \nat, 370, 194 

\bibitem[Ibata et al.(2001)]{Ibata2001} Ibata, R., Lewis, G.~F., Irwin, M., Totten, E., \& Quinn, T.\ 2001, \apj, 551, 294 

\bibitem[Ibata et al.(2013)]{Ibata2013} Ibata, R.~A., Lewis, G.~F., Conn, A.~R., et al.\ 2013, \nat, 493, 62 

\bibitem[Ibata et al.(2014)]{Ibata2014} Ibata, N.~G., Ibata, R.~A., Famaey, B., \& Lewis, G.~F.\ 2014, \nat, 511, 563 

\bibitem[Irwin et al.(1990)]{Irwin1990} Irwin, M.~J., Bunclark, P.~S., Bridgeland, M.~T., \& McMahon, R.~G.\ 1990, \mnras, 244, 16P 

\bibitem[Kroupa et al.(2005)]{Kroupa2005} Kroupa, P., Theis, C., \& Boily, C.~M.\ 2005, \aap, 431, 517 

\bibitem[Kroupa et al.(2010)]{Kroupa2010} Kroupa, P., Famaey, B., de Boer, K.~S., et al.\ 2010, \aap, 523, A32 

\bibitem[Kunkel \& Demers(1976)]{KunkelDemers1976} Kunkel, W.~E., \& Demers, S.\ 1976, The Galaxy and the Local Group, 182, 241 

\bibitem[K{\"u}pper et al.(2015)]{Kuepper2015} K{\"u}pper, A.~H.~W., Balbinot, E., Bonaca, A., et al.\ 2015, \apj, 803, 80 

\bibitem[Law \& Majewski(2010)]{LawMajewski2010} Law, D.~R., \& Majewski, S.~R.\ 2010, \apj, 714, 229 

\bibitem[Lelli et al.(2016)]{Lelli2016} Lelli, F., McGaugh, S.~S., Schombert, J.~M., \& Pawlowski, M.~S.\ 2016, arXiv:1610.08981 

\bibitem[Libeskind et al.(2009)]{Libeskind2009} Libeskind, N.~I., Frenk, C.~S., Cole, S., Jenkins, A., \& Helly, J.~C.\ 2009, \mnras, 399, 550 

\bibitem[Libeskind et al.(2015)]{Libeskind2015} Libeskind, N.~I., Hoffman, Y., Tully, R.~B., et al.\ 2015, \mnras, 452, 1052 

\bibitem[Lipnicky \& Chakrabarti(2016)]{LipnickyChakrabarti2016} Lipnicky, A., \& Chakrabarti, S.\ 2016, arXiv:1612.07325 

\bibitem[Lynden-Bell(1976)]{LyndenBell1976} Lynden-Bell, D.\ 1976, \mnras, 174, 695 

\bibitem[Maji et al.(2017a)]{Maji2017a} Maji, M., Zhu, Q., Marinacci, F., \& Li, Y.\ 2017a, arXiv:1702.00485 

\bibitem[Maji et al.(2017b)]{Maji2017b} Maji, M., Zhu, Q., Marinacci, F., \& Li, Y.\ 2017b, arXiv:1702.00497 

\bibitem[Metz et al.(2009)]{Metz2009} Metz, M., Kroupa, P., \& Jerjen, H.\ 2009, \mnras, 394, 2223 

\bibitem[M{\"u}ller et al.(2016)]{Mueller2016} M{\"u}ller, O., Jerjen, H., Pawlowski, M.~S., \& Binggeli, B.\ 2016, \aap, 595, A119 

\bibitem[Pawlowski et al.(2012)]{Pawlowski2012} Pawlowski, M.~S., Pflamm-Altenburg, J., \& Kroupa, P.\ 2012, \mnras, 423, 1109 

\bibitem[Pawlowski \& Kroupa(2013)]{PawlowskiKroupa2013} Pawlowski, M.~S., \& Kroupa, P.\ 2013, \mnras, 435, 2116 

\bibitem[Pawlowski et al.(2014)]{Pawlowski2014} Pawlowski, M.~S., Famaey, B., Jerjen, H., et al.\ 2014, \mnras, 442, 2362 

\bibitem[Pawlowski \& McGaugh(2014)]{PawlowskiMcGaugh2014} Pawlowski, M.~S., \& McGaugh, S.~S.\ 2014, \apjl, 789, L24 

\bibitem[Pawlowski et al.(2015)]{Pawlowski2015} Pawlowski, M.~S., Famaey, B., Merritt, D., \& Kroupa, P.\ 2015, \apj, 815, 19 

\bibitem[Pawlowski(2016)]{Pawlowski2016} Pawlowski, M.~S.\ 2016, \mnras, 456, 448 

\bibitem[Piatek et al.(2016)]{Piatek2016} Piatek, S., Pryor, C., \& Olszewski, E.~W.\ 2016, \aj, 152, 166 

\bibitem[Pryor et al.(2015)]{Pryor2015} Pryor, C., Piatek, S., \& Olszewski, E.~W.\ 2015, \aj, 149, 42 

\bibitem[Sawala et al.(2014)]{Sawala2014} Sawala, T., Frenk, C.~S., Fattahi, A., et al.\ 2014, arXiv:1412.2748 

\bibitem[Smith et al.(2016)]{Smith2016} Smith, R., Duc, P.~A., Bournaud, F., \& Yi, S.~K.\ 2016, \apj, 818, 11 

\bibitem[Sohn et al.(2017)]{Sohn2017} Sohn, S.~T., Patel, E., Besla, G., et al.\ 2017, arXiv:1707.02593 

\bibitem[The Dark Energy Survey Collaboration(2005)]{DESpaper} The Dark Energy Survey Collaboration 2005, arXiv:astro-ph/0510346 

\bibitem[Tollerud et al.(2008)]{Tollerud2008} Tollerud, E.~J., Bullock, J.~S., Strigari, L.~E., \& Willman, B.\ 2008, \apj, 688, 277-289 

\bibitem[Tully et al.(2015)]{Tully2015} Tully, R.~B., Libeskind, N.~I., Karachentsev, I.~D., et al.\ 2015, \apjl, 802, L25 

\bibitem[Yang et al.(2014)]{Yang2014} Yang, Y., Hammer, F., Fouquet, S., et al.\ 2014, \mnras, 442, 2419 

\bibitem[York et al.(2000)]{SDSSpaper} York, D.~G., Adelman, J., Anderson, J.~E., Jr., et al.\ 2000, \aj, 120, 1579 

\bibitem[Wang et al.(2013)]{Wang2013} Wang, J., Frenk, C.~S., \& Cooper, A.~P.\ 2013, \mnras, 429, 1502 

\bibitem[Watkins et al.(2010)]{Watkins2010} Watkins, L.~L., Evans, N.~W., \& An, J.~H.\ 2010, \mnras, 406, 264 

\end{thebibliography}
\end{document}